\documentclass[graybox]{svmult}

\usepackage{mathptmx}       
\usepackage{helvet}         
\usepackage{courier}        
\usepackage{type1cm}        
\usepackage{makeidx}         
\usepackage{graphicx}        
\usepackage{multicol}        
\usepackage[bottom]{footmisc}

\makeindex             

\newcommand{\oversim}[2]{\protect{\mbox{\lower0.5ex\vbox{%
   \baselineskip=0pt\lineskip=0.2ex
   \ialign{$\mathsurround=0pt #1\hfil##\hfil$\crcr#2\crcr\sim\crcr}}}}} 
\newcommand{\simless} {\mbox{$\,\mathrel{\mathpalette\oversim<}\,$}} 
\begin{document}

\title*{Recent advances on IMF research}
\author{Pavel Kroupa}
\institute{Pavel Kroupa \at Argelander-Institute for Astronomy,
  University of Bonn, Auf dem H\"ugel 71, 53121 Bonn, Germany, \email{pavel@astro.uni-bonn.de}
}
%
%
\maketitle

\abstract{ Here I discuss recent work on brown dwarfs, massive stars
  and the IMF in general, which are areas of research to which Anthony
  Whitworth has been contributing major work. The stellar IMF can be
  well described by an invariant two-part power law in present-day
  star-formation events (SFevs) within the Local Group of galaxies. It
  is nearly identical in shape to the pre-stellar core mass function
  \cite{Andre_etal10}. The majority of brown dwarfs follow a separate
  IMF. Evidence from globular clusters and ultra-compact dwarf
  galaxies has emerged that IMFs may have been top heavy depending on
  the star-formation rate density \cite{Marks_etal12}. The IGIMF then
  ranges from bottom heavy at low galaxy-wide star formation rates to
  being top-heavy in galaxy-scale star bursts.  }

\section{Introduction}
\label{sec:introd}

The stellar IMF is the distribution function of stellar masses, $m$,
formed together in one star-formation event (SFev) which can be
characterised by a spatial scale of up to about a pc and a stellar
mass $M_{\rm ecl}$. Various forms of distribution functions describing
the observationally derived IMF have been proposed
(e.g. \cite{CW12}). According to the recent Herschel results
(e.g. \cite{Andre_etal10, Andre_etal11}) the SFevs occur along thin (width of about
0.1~pc) filaments in the molecular clouds when the mass per unit
length surpasses about $15 \, M_\odot/$pc. The SFevs are deeply
embedded and the star formation efficiency is $\epsilon\approx
0.3-0.4$, such that about 60--70~per~cent of the residual gas is
blown out from them leaving the stellar population of mass
$few\,M_\odot \simless M_{\rm ecl}$ largely unbound. Taking the energy
distribution of binary populations in observed star clusters to limit
the largest density the cluster was allowed to have when it was a SFev
(too many binaries would be burned at too high densities), \cite{MK12}
inferred a radius-mass relation, $R_{\rm ecl}=0.10\,(M_{\rm
  ecl}/M_\odot)^{0.13}$, which extracts the same length scale. This
suggests that the universal initial binary distribution function
\cite{Kroupa_etal12} deduced from the many observations appears to be
a good representation of reality.

Remarkable progress has been achieved in constraining the form of the
IMF and its variability. This progress has largely been driven by
observational studies, but theoretical advances have also been
many. Here a brief review if provided of the recent issues concerning
the IMF, some of which are hotly debated if not poorly
understood. Further details and references are to be found in the
reviews by \cite{Chabrier03}, \cite{Bonnell_etal07},
\cite{Bastian_etal10} and \cite{Kroupa_etal12}.

\section{Universality of the IMF}
\label{sec:IMFuniv}

As is evident from the reviews mentioned above, a consensus appears to
have emerged in the community that the stellar IMF is largely
invariant for star formation conditions as are found throughout the
Local Group of galaxies at the present time. The form of this
universal or canonical IMF is most simply described by a two-part
power-law, $\alpha_1\approx 1.3, 0.07<m/M_\odot \simless 0.5$ and
$\alpha_2=2.3, 0.5 \simless m/M_\odot$ (the "Massey-Salpeter"
power-law index). This form can also be approximated by a log-normal
for $m<1\,M_\odot$ and the same power-law part for $m>1\,M_\odot$
\cite{Kroupa_etal12} but leads to a mathematically more complex object
without the gain of physical reality. Concerning the origin of the
IMF, \cite{Andre_etal10} note the remarkable similarity between the
pre-stellar core mass function and the stellar IMF, "suggesting a
$\sim$ one-to-one correspondence between core mass and star/system
mass with $M_{*,{\rm sys}} = \epsilon\,M_{\rm core}$ and $\epsilon
\approx 0.4$ in Aquila."

As will be seen below, the evidence that the brown dwarf IMF forms a
separate distribution function which is not a continuous extension of
the stellar IMF makes use of a log-normal form at low masses less
attractive.

\section{The brown dwarf issue}
\label{sec:BD}

It has been known for some time that brown dwarfs (BDs) are unlikely to
from form direct gravitational collapse in a molecular cloud such that
the observationally deduced mass function contains a significant
surplus of brown dwarfs \cite{PN02, Padoan_etal07, Andersen_etal11,
  Hennebelle12}.\footnote{An interesting sociological effect appears
  to have emerged in that authors claim good agreement with the
  (observed) Chabrier IMF but scrutiny of the published work shows
  consistently significant disagreement.} The reason is that the
distribution of density maxima in a cold but turbulent molecular cloud
has very few peaks which can collapse through eigengravity at the mass
scale of a BD such that not much further material is
accreted. Although \cite{Whitworth_etal07, Whitworth_etal10} argue that
BDs form a continuous extension of the stellar distribution, the
observational and theoretical evidence they provide strongly suggests
that BDs and stars have different properties in terms of their
pairing. \cite{Kroupa_etal03} have tested the hypothesis that BDs and
stars follow the exact same distribution functions and exclude this
hypothesis with very high confidence. The various flavours of BDs that
can in principle arise (collisional, photo-evaporated, ejected
embryos) have been discussed \cite{KB03} with the result that in the
present-day star-forming conditions mostly the ejected embryo flavour
dominates. The original suggestion of this scenario has been updated
by \cite{Stamatellos_etal07, Thies_etal10, BV12} by the argument that
the gravitationally pre-processed material in outer accretion disks is
able to cool sufficiently rapidly upon compression to allow direct
gravitational collapse at the BD mass scale.  The resulting IMF of BDs
compares remarkably well with the observationally deduced BD IMF
($\alpha_0\approx 0.3$). The resulting binary properties of BDs are
also accounted for naturally \cite{Thies_etal10}.

The BD IMF is thus a nearly flat power-law from the opacity limit for
fragmentation ($m_l \approx 0.01\,M_\odot$) to an upper limit which
transgresses the hydrogen burning limit. In principle, arbitrarily
massive "BDs" can form in very massive disks around massive stars such
that here the origin of stars vs BDs becomes blurred. Because massive
stars are exceedingly rare the stellar population formed through this
disk-fragmentation channel is negligible in comparison to the "normal"
stellar population which results from direct molecular cloud
fragmentation.

Thus in order to correctly account for a stellar population with BDs
most of the BD population must be added in terms of a separate
distribution function, as is also the case for planets which follow
their own mass distribution. The BD IMF can be expressed as a nearly
flat power-law with a continuous log-normal extension from the stellar
regime being ruled out.

\section{Variation of the IMF}
\label{sec:IMFvar}

A hint at a possible variation of the IMF in the MW has emerged due to
present-day star-formation events possibly producing more low-mass
stars than previously. This has been quantified as a metallicity
dependence, $\alpha_1\approx1.3 + 0.5[{\rm Fe/H}]$
\cite{Kroupa_etal12}. From the study of massive elliptical (E)
galaxies, it has emerged that the IMF must have been significantly
bottom heavy.  \cite{Cenarro_etal03} inferred $\alpha= 3.41+2.78[{\rm
  Fe/H}] - 3.79[{\rm Fe/H}]^2$ (for $0.1\simless m/M_\odot \simless
100$, although not explicitly stated in the paper) and a more recent
analysis by \cite{vDC11} also suggests an increasingly bottom heavy
IMF with increasingly massive E galaxies. This may be related to the
postulated cooling-flow-accretion population of low-mass stars
\cite{KG94}. A consistency check by Smith \& Lucey
(\cite{Smith_Lucey13} using gravitational lensing appears to exclude a
bottom-heavy IMF though.

\section{The massive end of the IMF}
\label{sec:IMFmassive}

\cite{Whitworth_etal94} had already suggested that massive stars may
preferentially form in shocked gas. As reviewed in
\cite{Kroupa_etal12} there has been much observational evidence for
top-heavy IMFs in star-bursts.  As these are observationally
unresolved, this evidence was indirect and largely
ignored. Observations of the assembly of the stellar population over
cosmological epoch have also been pointing to top-heavy IMFs in the
past, as otherwise there would be more low-mass stars locally than are
observed. Three independent more-direct lines of evidence for the IMF
becoming top-heavy with star-formation rate density have recently
emerged:

First: It is well known that ultra-compact dwarf galaxies (UCDs),
which have a mass scale of $10^6-10^8\,M_\odot$, have larger dynamical
mass-to-light (M/L) ratios than normal stellar populations. This is
unlikely due to exotic dark matter as the phase-space available in
UCDs would not accommodate significant amounts of dark
matter. Instead, a top-heavy IMF would have lead to an overabundance
of stellar remnants in UCDs which would enhance their dynamical M/L
ratios. Thus, the variation of the required $\alpha_3, m>1\,M_\odot,$
can be sought to explain the dynamical M/L ratia
\cite{Dabringhausen_etal09}.

Secondly: UCDs have an overabundance of low-mass X-ray bright sources
(LMXBs). In globular clusters (GCs), LMXBs are known to be formed from
the dynamical capture of stars by stellar remnants mostly in the core
of the GCs. As the star evolves the remnant accretes part of the
star's envelope thus becoming detectable with X-rays. The LMXB
population is constantly depopulating and needs to be replenished by
new capture events. Indeed, the theoretically expected scaling of the
fraction of GCs with LMXB sources with GC mass is nicely consistent
with the observed data assuming an invariant stellar MF. Applying the
same theory to UCDs uncovers a break-down of this agreement as the
UCDs have a surplus of LMXB sources. By adding stellar remnants
through a top-heavy IMF when the UCDs were born, i.e. by allowing
$\alpha_3$ to vary with UCD birth mass, consistency with the data can
be sought \cite{Dabringhausen_etal12}.

Thirdly: Low-concentration GCs have been found by de Marchi et
al. (2007) to be depleted in low mass stars while high-concentration
GCs have a normal MF. This is contrary to the energy-equipartition
driven depopulation of low mass stars because more concentrated
clusters ought to have lost more low mass stars. It is also not
consistent with any known theory of star formation, because the
low-concentration clusters typically have a higher metallicity which
would, if anything, imply a surplus of low-mass stars. The currently
only physically plausible explanation is to suppose that the young GCs
formed compact and mass segregated and that the expulsion of residual
gas unbound a part of the low-mass stellar population. By constraining
the necessary expansion of the proto-GCs (i.e. SFevs), correlations
between metallicity, $\alpha_3$ and tidal field strength emerge which
constrain the very early sequence of events that formed the Milky Way
as well as the dependency of $\alpha_3$ on density and metallicity of
the SFev \cite{Marks_etal12}.

Putting this all together, a consistent variation of $\alpha_3$ with
density and metallicity of the SFevs emerges: for $m>1\,M_\odot, x \ge
-0.89$: $\alpha_3 = -0.41\,x + 1.94$ with $x=-0.14[{\rm Fe/H}] + 0.99
{\rm log}_{10}(\rho_6)$, where $\rho_6 = \rho / (10^6 M_\odot {\rm
  pc}^{-3} )$ and $\rho$ is the density in $M_\odot/{\rm pc}^3$. 

Thus, SFevs at a star-formation rate density $SFRD<0.1\,M_\odot/({\rm
  pc}^3\,{\rm Myr})$ can be assumed to have an invariant IMF with
$\alpha_3=\alpha_2$ (subject to the possible variation with
metallicity discussed above), while SFevs with larger SFRDs tend
towards top-heavy IMFs whereby the trend is enhanced at
lower-metallicities.

\section{Massive stars and the IGIMF}
\label{sec:IGIMF}

The formation of massive stars is notoriously difficult to study
because they are rare and deeply embedded.  Thus, much fiction can be
associated with the formation of massive stars and the only well-posed
approach to ascertain a hypothesis is to test its consequences against
data taking care to note that by showing one hypothesis to work does
not exclude another hypothesis.

There are two major competing hypothesis: 

According to the one hypothesis the IMF may be taken to be a
probability distribution function such that the stellar ensemble in a
whole galaxy is always a random draw from the stellar IMF. This allows
massive stars to form in isolation as rare events.

The other hypothesis is related to optimal sampling
\cite{Kroupa_etal12} according to which the stellar IMF is a
distribution function which scales with $M_{\rm ecl}$ such that the
most massive star, $m_{\rm max}$, in the SFev follows a $m_{\rm
  max}(M_{\rm ecl})$ relation \cite{WKB10}.  The total star-formation
rate (SFR) of a galaxy follows from all its SFevs, such that a large
SFR implies SFevs that reach to large masses and thus to large SFRDs
which then imply top-heavy IMFs in these. As a consequence the IMF of
a whole galaxy (the "integrated IMF" $=$ IGIMF) is steeper (larger
$\alpha_3$), or flatter (i.e. top-heavy) than Massey-Salpeter,
depending on its SFR. The implications for the astrophysics of
galaxies as well as for cosmology are major.  

The vast quantity of data are consistent with the latter theory, and
most data can most simply and naturally be explained within the IGIMF
framework \cite{Pf11, W11}. A counter-argument against the IGIMF
theory often put up, namely that evidence exists that massive stars
can form in isolation, is countered by the observed fraction of massive
stars deemed to have formed in isolation being smaller than the
fraction of apparently isolated massive stars if all massive stars in
fact do form in embedded clusters, and by virtually all
best-candidates for isolated massive star formation having been shown
to be most likely stemming from clusters \cite{Gvaramadze_etal12}.

%
%
%

\end{document}